# MODELING OF T-SHAPED MICROCANTILEVER RESONATORS

*Margarita Narducci, Eduard Figueras, Isabel Gràcia, Luis Fonseca, Joaquin Santander, Carles Cané*

Centro Nacional de Microelectrónica de Barcelona, CNM-IMB (CSIC), Campus UAB, 08193 Bellaterra, Barcelona, Spain. Phone: +34935947700, Fax: +34935801496
Email: margarita.narducci@cnm.es

## ABSTRACT

The extensive research and development of micromechanical resonators is trying to allow the use of these devices for highly sensitive applications. Microcantilevers are some of the simplest MEMS structure and had been proved to be a good platform due to its excellent mechanical properties. A cantilever working in dynamic mode, adjust its resonance frequency depending on changes in both, the spring constant ($k$) and mass ($m$) of the resonator. The aim of this work was to model a cantilever structure to determine the optimal dimensions in which the resonance frequency would be a function dominated by mass changes and not stiffness changes. In order to validate the model a set of microcantilevers were fabricated and characterized.

## 1. INTRODUCTION

Silicon microcantilevers have been increasingly used for sensor applications, due mainly to its high sensitivity. This type of sensor can be classified into two kinds, depending on the mode of operation, static and dynamic. The static cantilever detects deformations and the dynamic detects resonant frequency shift. The present work is focused on dynamic cantilevers.

The resonance frequency on a cantilever working in dynamic mode is given by equation (1).

$$f = \frac{1}{2\pi}\sqrt{\frac{k}{m_{eff}}} \quad (1)$$

Where $k$ is the spring constant and $m_{eff}$ is the effective or dynamic mass. As show equation (1), any change in $m_{eff}$ and $k$ will change the frequency of the cantilever. However a varying spring constant is one variable that is usually not taken into consideration in studies dealing with cantilever sensors [2].

This paper proposes a model of a T-shaped cantilever in which the change on the spring constant caused by an extra mass has been included in the calculations. This model allow to determine the optimal geometry dimensions of the structure in order to make the resonance frequency dominated by mass changes instead of stiffness and therefore improve the sensitivity of the cantilever resonator.

## 2. CANTILEVER DESIGN

The cantilevers were designed to resonate in flexural mode perpendicular to the substrate and were driven to reach their mechanical resonance by a ceramic-insulated multilayer piezoactuator glued at the backside and the resonance frequency was measured by four piezoresistors in a Wheatstone bridge configuration. The resonators were designed as three beams that are hold together by means of an extra rectangular mass, as can be seen in figure 1.

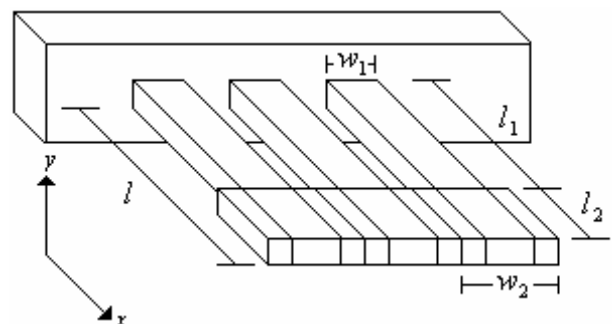

Figure 1. Diagram of a T-shaped cantilever.

Where $l_1$, $w_1$ and $h$ are, respectively, the length, width and thickness of each beam, $l_2$ and $w_2$ the length and width of each extra mass.

*Margarita Narducci, Eduard Figueras, Isabel Gràcia, Luis Fonseca, Joaquin Santander, Carles Cané*
*Modeling of T-Shaped Microcantilever Resonators*

The resonance frequency of each single cantilever after the attachment of the extra mass is given by the next equation [15]:

$$f = \frac{1}{2\pi}\sqrt{\frac{k+\Delta k}{0,24m_1 + m_2}} \quad (2)$$

Where $k+\Delta k$ is the spring constant after the mass is added, $0,24m_1$ is the effective mass of the cantilever beam and $m_2$ is the added mass.

In order to calculate the spring constant after the attachment of the mass ($k+\Delta k$), the general equation of the elastic curve was used:

$$\frac{d^2y}{dx^2} = \frac{-F}{EI}(x-L) \quad (3)$$

Where $F$ is the applied force to the structure, $E$ is the Young modulus of the material and $I$ is the moment of inertia. The model was simplified assuming that the three beams oscillate in the first mode and at the same time, therefore the equation (3) was applied to one single beam and its attached mass. As the equation (3) depend on $x$, it was necessary to split the cantilever into two regions. The first region lay between $0 \leq x \leq l_1$, a mass free cantilever, and the second one between $l_1 \leq x \leq l_2$, the extra mass [5]. That is:

$$\frac{d^2y}{dx^2} = \begin{cases} \frac{d^2y_1}{dx^2} = \frac{-F}{EI_1}(x-(l_1+l_2)); x \leq l_1 \\ \frac{d^2y_2}{dx^2} = \frac{-F}{EI_2}(x-(l_1+l_2)); x \geq l_1 \end{cases} \quad (4)$$

Where $I$ is the area moment of inertia of a rectangular beam and can be expressed as:

$$I_1 = \frac{w_1 h^3}{12}; \quad I_2 = \frac{w_2 h^3}{12} \quad (5)$$

The equation (4) was integrated two times in order to find the deflection $y(x)$, see equation (6):

$$y(x) = \begin{cases} y_1 = \frac{-F}{EI_1}(\frac{x^3}{6} - (l_1+l_2)\frac{x^2}{2} + ax + b); x \leq l_1 \\ y_2 = \frac{-F}{EI_2}(\frac{x^3}{6} - (l_1+l_2)\frac{x^2}{2} + cx + d); x \geq l_1 \end{cases} \quad (6)$$

Subjected to:

$$y_1(0) = 0; y_1'(0) = 0; y_1''(0) = 0$$
$$y_1(l_1) = y_2(l_1); y_1'(l_1) = y_2'(l_1) \quad (7)$$

Using the cantilever boundary conditions, the integration constants were calculated:

$$a = 0$$
$$b = 0$$
$$c = -\frac{(-l_1^2 w_1 - 2l_1 l_2 w_1 + l_1^2 w_2 + 2l_1 l_2 w_2)}{2w_1} \quad (8)$$
$$d = -\frac{(l_1^3 w_1 + 3l_1^2 l_2 w_1 - l_1^3 w_2 - 3l_1^2 l_2 w_2)}{6w_1}$$

Substituting (5) and (8) into (6), the equation for the deflection $y(x)$ was obtained. Then it is necessary to evaluate the solution in $x = l_1 + l_2$ to determine $y_{max}$, that is:

$$y_{max} = \frac{4F}{Eh^3} \times \left(\frac{l_2^3}{w_2} + \frac{l_1^3}{w_1} + \frac{3l_1^2 l_2}{w_1} + \frac{3l_1 l_2^2}{w_1}\right) \quad (9)$$

Next, to calculate $k+\Delta k$, the general equation for the spring constant was used:

$$k = \frac{F}{y_{max}} \quad (10)$$

Where $F$ is the force acting on the spring, and $y_{max}$ is the maximum displacement of the spring. Replacing (9) into (10) and simplifying: the new spring constant can be shown to be:

$$k + \Delta k = \frac{Eh^3}{4} \times \frac{w_1 w_2}{(l_2^3 w_1 + l_1^3 w_2 + 3l_1^2 l_2 w_2 + 3l_1 l_2^2 w_2)} \quad (11)$$

If the corresponding area to the second region of the cantilever would be eliminated ($l_2 = 0$ and $w_2 = 0$), then the equation (11) would become into the well known equation of the spring constant for a simple rectangular cantilever [2, 5]:

$$k = \frac{Eh^3}{4} \times \frac{w_1}{l_1^3} \quad (12)$$

### 3. FABRICATION PROCESS

The fabrication process is illustrated in Figure 2. The resonators were fabricated following a 7-mask process, starting with an N-type SOI double side polish wafer. The silicon layer is 15µm thick over a 2µm buried oxide and 450µm of bulk silicon. The process starts with a 180Å grown dry oxide layer and then a 1175Å silicon nitride layer is deposited (Figure 2a). The first level mask is used to define the active zone on the frontside, and then a 10600Å field oxide is grown. After that, with the second mask the backside window is defined (Figure 2b). Next, the silicon nitride on the frontside is removed and a Boron implantation of $1\times 10^{15}$ cm$^{-2}$ and 50eV is performed to define the resistivity of the Wheatstone bridge resistors, subsequently trough the third mask a new implantation fix the resistivity of contacts and heaters. Then a 1.3µm BPTEOS oxide is deposited (Figure 2c). The fourth mask is used to open contacts. Aluminum is deposited and patterned using the fifth mask (Figure 2d) to define metal connections and bonding pads. Afterwards a 0.4µm PECVD oxide and 0.4µm PECVD nitride are deposited (passivation layer) and patterned with the sixth mask (Figure 2e). The seventh mask is used to define the motif on the frontside, after that using the nitride mask on the backside the silicon substrate is etched using a KOH bath. After that, the 15µm membrane is etched by reactive ion etching and the cantilever is released (Figure 2f). Finally the whole structure is glued to a ceramic-insulated multilayer piezoactuator. The figure 3 shows a photograph of one fabricated cantilever.

Two chips were fabricated. One chip (chip 1) contains structures of dimensions in the range of 400×300µm2 and the other (chip 2) contains structures of dimensions in the range of 200×150µm2. The dimensions of the fabricated cantilevers are shown in the table 1.

|   | $l$ * (µm) | $w_1$ (µm) | $l_2$ (µm) | $w_2$ (µm) |
|---|---|---|---|---|
| 1 | 400 | 64 | 350 | 100 |
| 2 | 400 | 64 | 300 | 100 |
| 3 | 400 | 64 | 200 | 100 |
| 4 | 400 | 64 | 70 | 100 |
| 5 | 200 | 32 | 175 | 50 |
| 6 | 200 | 32 | 150 | 50 |
| 7 | 200 | 32 | 100 | 50 |
| 8 | 200 | 32 | 30 | 50 |

Table 1. Dimensions of the cantilevers fabricated. For all structures the thickness $h$ is 15µm. *where $l = l_1 + l_2$

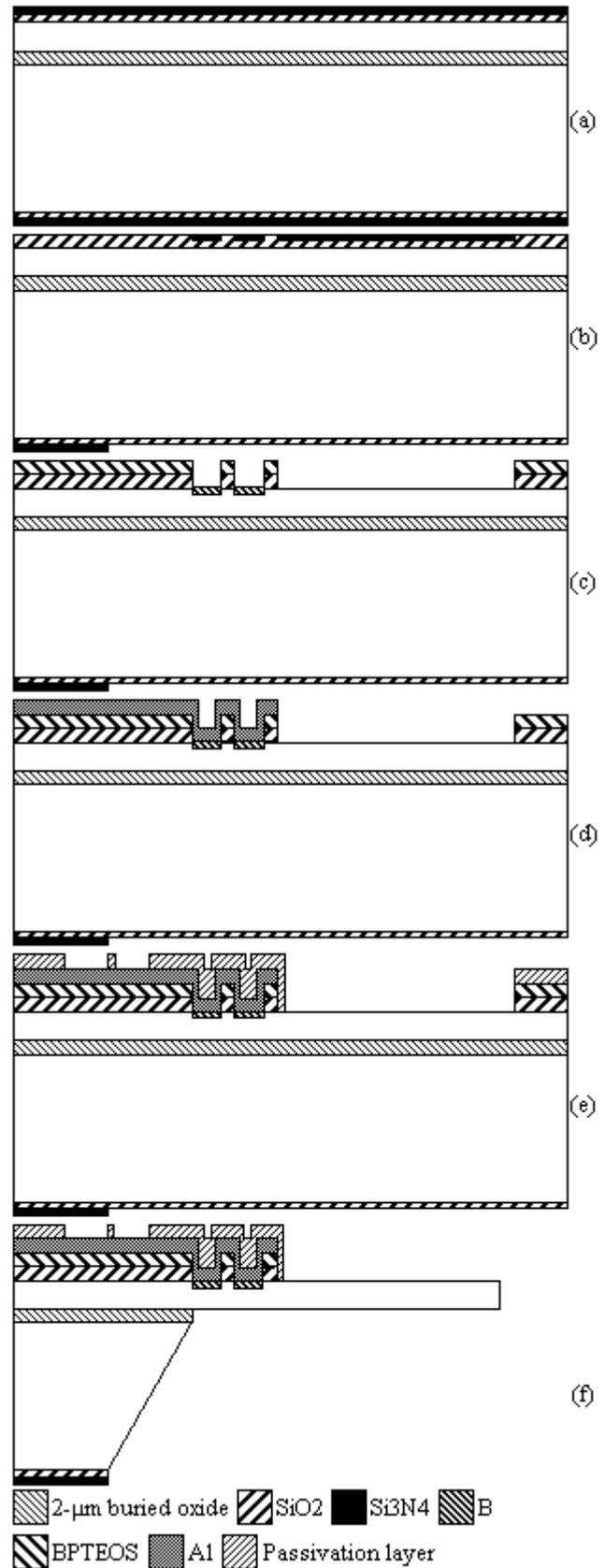

Figure 2. Fabrication process flow.

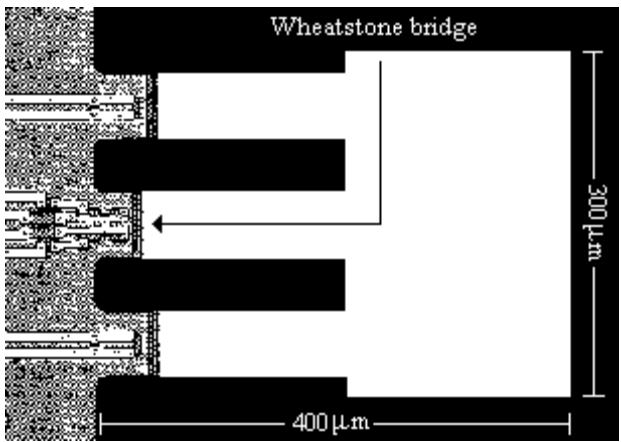

Figure 3. Photograph of a fabricated cantilever.

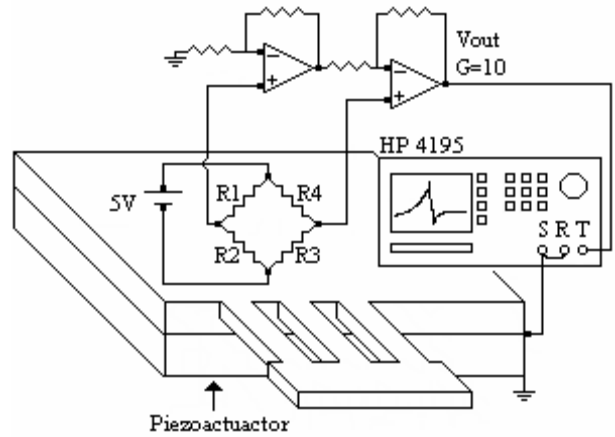

Figure 4. Experimental setup to perform the measurements.

### 4. DEVICE CHARACTERIZATION

To perform the characterization of the cantilevers, first the piezoactuator was connected to the source of the network analyzer (HP 4195) and the detector (Wheatstone bridge) to a voltage amplifier (G=10) which in turn was connected to the input of the HP 4195. The instrumental setup is shown in figure 4.

Using this experimental setup was possible to obtain the magnitude and phase response and therefore measure resonance frequency ($f_r$) and quality factor ($Q$). The measurements were performed with eight (8) samples of each cantilever. The average measured values are contained in table 2. The estimated measurement error was calculated to be $f_{r\_ave} \pm 1\%$ and $Q_{ave} \pm 15\%$.

|   | Resonance frequency ($f_r$) in KHz | Quality factor ($Q$) |
|---|---|---|
| 1 | 98  | 680  |
| 2 | 92  | 870  |
| 3 | 89  | 810  |
| 4 | 97  | 820  |
| 5 | 354 | 950  |
| 6 | 334 | 1200 |
| 7 | 324 | 1050 |
| 8 | 400 | 940  |

Table 2. Results for the experimental measurements.

In order to compare experimental values with the theoretical model (equation 2), the relationship between $f_r$ and extra mass is illustrated in figure 5.

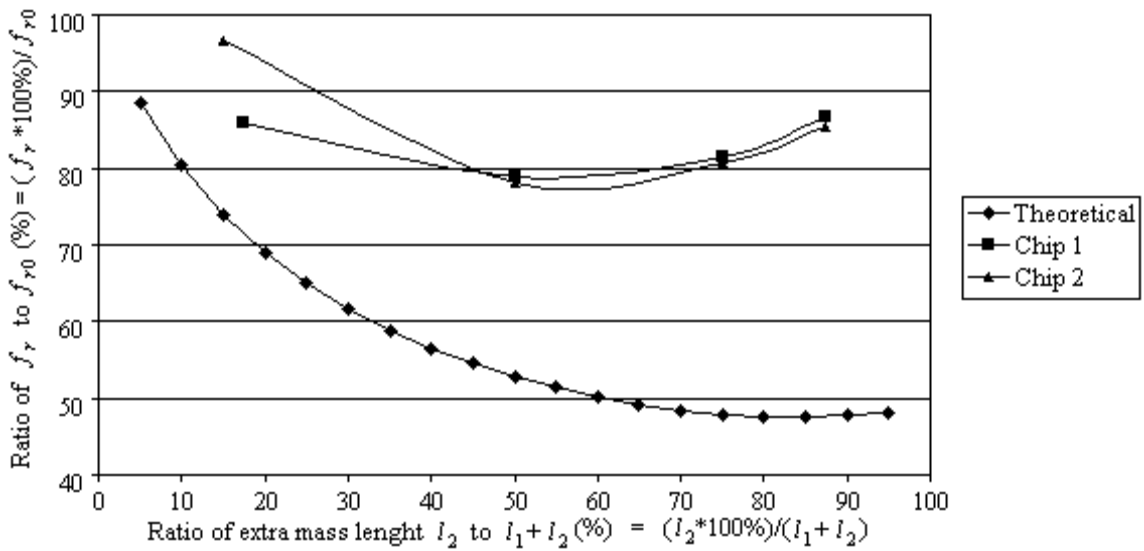

Figure 5. Plot of resonance frequency vs. extra mass length

*Margarita Narducci, Eduard Figueras, Isabel Gràcia, Luis Fonseca, Joaquin Santander, Carles Cané*
*Modeling of T-Shaped Microcantilever Resonators*

As can be seen from the theoretical trace in figure 5 there are two kinds of behaviors, the first one lay between $0 \leq l_2 \leq 0.8l$ and it is dominated by changes on the resonator extra mass (more mass, lower value of resonance frequency) and the second one $0.8l \leq l_2 \leq l$ dominated by changes on the spring constant (more mass, slightly higher value of resonance frequency). Whereas in the trace of the experimental measurements (chip 1 and chip 2), the two behaviors are delimited by the 60% of $l$. This difference among the theoretical and the experimental trace could be attribute to an error on estimating the effective mass owing to an increment of it due to the extra mass. In consequence, the effective mass of this cantilever should be recalculated. With the purpose of estimate a more accurate value of the effective mass, the cantilevers were simulated using Finite Element Analysis (FEA) with ANSYS. From the new values of $m_{eff}$ obtained with the FEA and the spring constant (computed with the equation 11) the resonance frequency was recalculate (equation 1). To compare these results with the experimental values, the relationship between $f_r$ and extra mass is illustrated in figure 6, which shows clearly a better agreement with the measured data.

## 5. CONCLUSIONS

In conclusion, this paper focuses on the design and modeling of a Silicon-based microcantilever resonator for highly sensitive applications. A T-shaped cantilever model for the resonant frequency had been calculated, finding a polynomial function with a minimum point marking the limit between the two types of behaviors previously mentioned. This minimum point for the added mass was measured to be in ~60% of the total length of the cantilever. So, attaching a rectangular mass with $l_2 < 0.6l$, shifts in frequency could be attribute to changes on the resonator mass. There was good agreement between predicted behavior for the resonant frequency and experimental values. Featuring high sensitivity, only to mass changes, this microcantilever resonator is promising to become a platform for sensor applications. Consequently, an extension of this work would be to deposit polymer on the cantilever in order to study the sensitivity of the resonator.

## 6. ACKNOWLEDGEMENTS

This work was supported by the CICYT-TIC-2002-0554-C03-02 project.

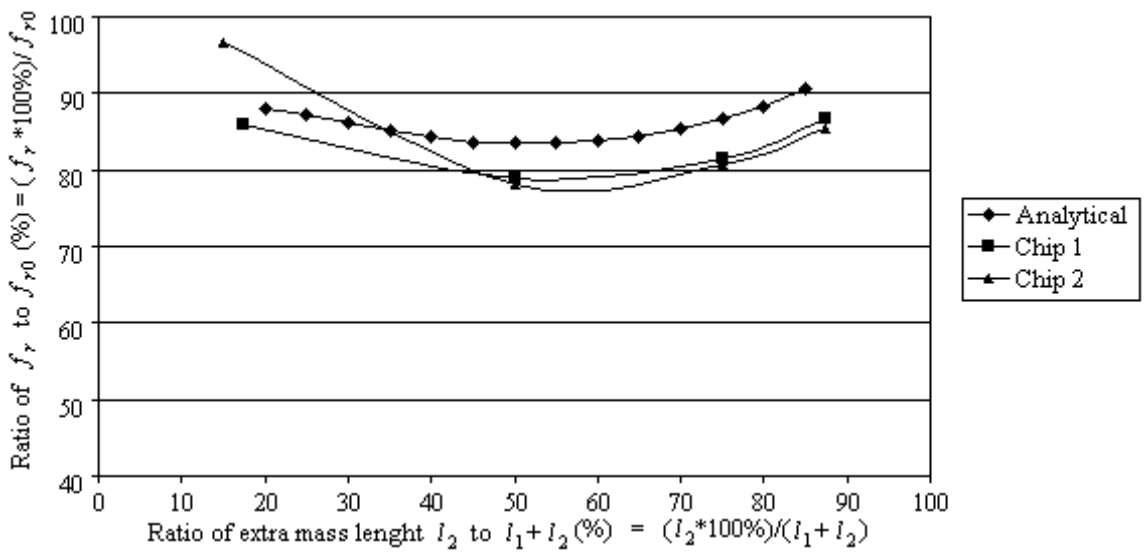

Figure 6. Plot of resonance frequency vs. extra mass length